\documentclass[
 aip,
 jcp,
amsfonts,
amsmath,
amssymb,
reprint,%
]{revtex4-1}

\usepackage[colorlinks,linkcolor=blue,urlcolor=blue,citecolor=blue]{hyperref}
\usepackage{bm}
\usepackage{graphicx}
\usepackage{subfig}

\def\bL{\mathbf{L}}
\def\bX{\mathbf{X}}
\def\bB{\mathbf{B}}
\def\bIx{\mathbf{I}_{1}}
\def\bIy{\mathbf{I}_{2}}
\def\bIz{\mathbf{I}_{3}}

\def\bLx{\mathbf{L}_{1}}
\def\bLy{\mathbf{L}_{2}}
\def\bLz{\mathbf{L}_{3}}
\def\bVx{\mathbf{V}_{1}}
\def\bVy{\mathbf{V}_{2}}
\def\bVz{\mathbf{V}_{3}}
\DeclareMathOperator{\erfc}{erfc}

\begin{document}

\title{Efficient real space formalism for hybrid density functionals}
\author{Xin Jing}
\affiliation{College of Engineering, Georgia Institute of Technology, Atlanta, GA 30332, USA}
\affiliation{College of Computing, Georgia Institute of Technology, Atlanta, GA 30332, USA}
\author{Phanish Suryanarayana}
\email[Email: ]{phanish.suryanarayana@ce.gatech.edu}
\affiliation{College of Engineering, Georgia Institute of Technology, Atlanta, GA 30332, USA}
\affiliation{College of Computing, Georgia Institute of Technology, Atlanta, GA 30332, USA}

\begin{abstract}
We present an efficient real space formalism for hybrid exchange-correlation functionals in generalized Kohn-Sham density functional theory (DFT). In particular, we develop an efficient representation for any function of the real space  finite-difference Laplacian matrix by leveraging its Kronecker product structure, thereby enabling the time to solution of associated linear systems to be highly competitive with the fast Fourier transform scheme,  while not imposing any restrictions on the boundary conditions. We implement this formalism for both the unscreened and range-separated variants of hybrid functionals. We verify its  accuracy and efficiency through comparisons with established planewave codes for isolated as well as bulk systems. In particular, we demonstrate up to an order-of-magnitude speedup in time to solution for the real space method. We also apply the framework to study the structure of liquid water using ab initio molecular dynamics, where we find good agreement with literature. Overall, the current formalism provides an avenue for efficient real space DFT calculations with hybrid density functionals. 
\end{abstract}

\maketitle

\section{Introduction}
Over the past few decades, electronic structure calculations based on Kohn-Sham density functional theory (DFT) \cite{kohn1965self, hohenberg1964inhomogeneous}  have become indispensable in materials and chemical sciences research due to the fundamental insights they provide and the predictive capabilities they offer. The widespread popularity of DFT, whose origins are in the first principles of quantum mechanics, can be attributed to its simplicity, generality, and high accuracy-to-cost ratio relative to other such ab initio methods \cite{becke2014dft, burke2012dft}. Though less expensive than wavefunction-based alternatives, Kohn-Sham calculations are still associated with significant cost, which restricts the range of systems that can be studied. These limitations become more pronounced for the choice of advanced exchange-correlation functionals, especially in ab initio molecular dynamics (AIMD) simulations, where tens or hundreds of thousands Kohn-Sham solutions  may be needed to study  the property/phenomena of interest \cite{burke2012dft}. 

The planewave method \cite{martin2020electronic} has been one of the most widely utilized solution strategies for the Kohn-Sham problem within the pseudopotential approximation \cite{VASP, CASTEP, ABINIT, Espresso, CPMD, DFT++, gygi2008architecture, valiev2010nwchem}. The underlying Fourier basis makes the method accurate, while also being  highly efficient on moderate computational resources through the use of well-optimized Fast Fourier Transforms (FFTs). However, the basis constrains the method to periodic boundary conditions and its global nature hinders scalability on parallel computing platforms. These limitations of the planewave method have spurred the development of alternative approaches  that utilize systematically improvable, localized representations  \cite{becke1989basis, chelikowsky1994finite, genovese2008daubechies, seitsonen1995real, white1989finite, iwata2010massively, tsuchida1995electronic, xu2018discrete, Phanish2011, Phanish2010, ONETEP, CONQUEST,MOTAMARRI2020106853, castro2006octopus, briggs1996real, fattebert1999finite, shimojo2001linear, ghosh2017sparc, arias1999wav, pask2005femeth, lin2012adaptive, xu2021sparc}.  Of these, real space finite-difference methods, where computational locality is maximized and Dirichlet as well as periodic/Bloch-periodic boundary conditions can naturally be accommodated, are perhaps the most mature and commonly employed to date. In particular, these methods can significantly outperform their planewave counterparts for local/semilocal exchange-correlation functionals, with increasing advantages as the number of processors is increased \cite{xu2021sparc, zhang2024sparc, GHOSH2017109}. Furthermore, they have been scaled to large systems  containing up to a million atoms \cite{gavini2022roadmap, fattebert2016modeling, dogan2023real}.

Hybrid density functionals are orbital-dependent exchange correlation functionals that are formulated within generalized Kohn-Sham DFT \cite{martin2020electronic}. These functionals, which are positioned on the fourth rung of Jacob's ladder, include a part of the nonlocal Hartree–Fock exact exchange energy, in addition to contributions from  local and semilocal exchange-correlation approximations.  Hybrid functionals   can be broadly classified into the unscreened and screened/range-separated variants, the former more popular for isolated systems such as molecules and clusters, while the latter more popular for condensed matter such as surfaces and 3D bulk systems. In particular, hybrid functionals offer better predictive capabilities than the local semilocal approximations for a range of properties such as the ionization potential, atomization energy, proton affinity, lattice constant, bulk modulus, heat of formation, spin magnetic moment, and bandgap \cite{gill1992investigation, becke1993new, Becke1993DensityfunctionalTI, becke1996density, adamo1999toward, adamo1999accurate, garza2016predicting}. 

Kohn-Sham calculations with hybrid functionals are substantially more expensive then their local/semilocal counterparts, by up to two orders-of-magnitude. This has motivated the development of a number of approaches to reduce the prefactor and/or scaling associated with such calculations  \cite{lin2016adaptively, gygi2009compact, damle2015compressed, damle2017scdm, damle2017computing, hu2017projected, hu2017interpolative, mountjoy2017exact, ko2020enabling, ko2021enabling, ko2023high}. In each of these approaches, the convolutions inherent to hybrid calculations are efficiently evaluated using FFTs. While these advancements can be integrated into the real-space method, the overall efficiency remains constrained by the need for employing iterative methods to solve the linear systems associated with the convolutional kernel  \cite{boffi2016efficient}. Furthermore, only  unscreened hybrid functionals can be implemented following this strategy \cite{boffi2016efficient, wu2009order}, whereby range-separated hybrid functionals have remained unexplored within the real space method. An alternate strategy to reduce the cost associated with the convolutions in real space is to employ a Tucker-tensor based decomposition, which allows for the 3D convolution to be written as a tensorial product of 1D convolutions \cite{subramanian2024tucker}. However, this approach is less straightforward and efficient compared to FFTs, and has currently only been demonstrated for isolated systems and unscreened hybrid functionals.

In this work, we present an efficient real space formalism for hybrid exchange-correlation functionals in generalized Kohn-Sham DFT. In particular, we develop an efficient representation for any function of the real-space finite-difference Laplacian matrix, leveraging its Kronecker product structure. Thereafter, associated linear systems can be solved with the efficiency of the FFT scheme, while accommodating Dirichlet as well as periodic/Bloch-periodic boundary conditions. We implement this formalism for both the unscreened and range-separated variants of hybrid functionals, and verify its accuracy and efficiency by comparing against established planewave codes for isolated as well as bulk systems. In particular, we demonstrate up to an order-of-magnitude speedup in time to solution for the real space method. We also apply the framework to study the structure of liquid water using AIMD, where we find good agreement with literature. 

The remainder of the manuscript is organized as follows. In Section~\ref{Sec:hybridDFT}, we provide some mathematical background on hybrid functionals. In Section~\ref{Sec:Formulation}, we discuss the Kronecker product based real space formalism for hybrid functionals. In Section~\ref{Sec:Implementation}, we discuss details of its implementation, and verify its accuracy and efficiency in Section~\ref{Sec:Results}.  Finally, we provide concluding remarks in Section~\ref{Sec:Conclusions}. 


\section{Hybrid density functionals} \label{Sec:hybridDFT}

We now briefly discuss the contribution of the exact exchange to the unscreened and range-separated variants of hybrid exchange-correlation functionals \cite{martin2020electronic}. 


\subsection{Unscreened} \label{Subsec:Unscreened}
In unscreened hybrid functionals, the exchange-correlation energy includes a certain fraction of the complete exact exchange energy, i.e., one that includes both the short- and long-range parts:
\begin{align}
    E_X &= -\frac{1}{2} \sum_{\sigma} \sum_{n\bm{k},m\bm{q}} w_{\bm{k}} w_{\bm{q}} g_{n\bm{k}}^{\sigma} g_{m\bm{q}}^{\sigma} \nonumber \\
    & \times \int  \int 
    \frac{ \psi_{n\bm{k}}^{\sigma\ast}(\bm{r}) \psi_{m\bm{q}}^{\sigma\ast}(\bm{r'}) \psi_{n\bm{k}}^{\sigma}(\bm{r'}) \psi_{m\bm{q}}^{\sigma}(\bm{r})
    }{|\bm{r}-\bm{r'}|} \, d\bm{r'}  d\bm{r} , \label{Eq:Ex}
\end{align}
where $\psi$ are the orbitals, $g$ are the occupations, and $w$ are the Brillouin zone weights, with the quantities being indexed by the spin $\sigma \in \{ \uparrow, \downarrow\}$,   Brillouin zone wavevectors $\bm k$ and $\bm q$, and the band numbers $m$ and $n$. The corresponding potential appearing in the Hamiltonian, referred to as the exact exchange operator, takes the form:
\begin{align}
    &V_{X}^{\sigma}\varphi_{n\bm k}^{\sigma}(\bm r)= -\sum_{m\bm q} w_{\bm q} g_{m\bm q}^{\sigma} \psi_{m\bm q}^{,\sigma}(\bm r)
    \int   \frac{\psi_{m\bm q}^{\sigma\ast}(\bm{r'}) \varphi_{n\bm k}^{\sigma}(\bm{r'}) }{|\bm{r}-\bm{r'}|}  d\bm{r'} \,, \label{Eq:Vx}
\end{align} 
where $\varphi$ is any given function, and $(.)^*$ represents the complex conjugate. Noting that the Coulomb kernel is the Green's function of the 3D Laplace operator (times $-1/4\pi$), the exact exchange operator in Eq.~\ref{Eq:Vx} and the corresponding energy in Eq.~\ref{Eq:Ex} can be rewritten as: 
\begin{align}
V_{X}^{\sigma}\varphi_{n\bm k}^{\sigma}(\bm r) & = -\sum_{m\bm q} w_{\bm q} g_{m\bm q}^{\sigma} \psi_{m\bm q}^{,\sigma}(\bm r)  \phi_{m\bm q n\bm k}^{\sigma} (\bm{r}) \,, \label{Eq:Vx:Poisson} \\
 E_X & = -\frac{1}{2} \sum_{\sigma} \sum_{n\bm{k},m\bm{q}} w_{\bm{k}} w_{\bm{q}} g_{n\bm{k}}^{\sigma} g_{m\bm{q}}^{\sigma}  \nonumber \\ 
& \hspace{6mm} \times \int \psi_{n\bm{k}}^{\sigma\ast}(\bm{r}) \phi_{m\bm q n\bm k}^{\sigma} (\mathbf{r})  \psi_{m\bm{q}}^{\sigma}(\bm{r}) dr \,, \label{Eq:Ex:Poisson} 
\end{align}
where $\phi$ corresponds to the solution of the Poisson equations: 
\begin{equation}\label{Eq:Poisson}
- \frac{1}{4 \pi} \nabla^2 \phi_{m\bm q n\bm k}^{\sigma} (\mathbf{r}) =  \psi_{m\bm q}^{\sigma\ast}(\bm{r}) \varphi_{n\bm k}^{\sigma}(\bm{r}) \,,
\end{equation}
subject to Bloch boundary conditions at the  wavevector $\bm k - \bm q$ in the directions that the system is extended, and  Dirichlet boundary conditions in the other directions, i.e., vacuum. Indeed, the expressions are applicable for isolated systems, i.e., systems with vacuum in all  directions, by setting the wavevectors $\bm k = \bm q=0$ and the Brillouin zone weight $\bm{w}_0=1$. 

\subsection{Range-separated} \label{Subsec:RangeSeparated}
In range-separated hybrid functionals, the exchange-correlation energy includes a certain fraction of only the short-range part of the exact exchange energy, with the range partitioning typically achieved  through the complementary error function ($\erfc$):
\begin{align}\label{Eq:ExSR}
    &E_{X,SR} = -\frac{1}{2} \sum_{\sigma} \sum_{n\bm{k},m\bm{q}} w_{\bm{k}} w_{\bm{q}} g_{n\bm{k}}^{\sigma} g_{m\bm{q}}^{\sigma} \nonumber \\
    & \times \int   \int
    \frac{ \psi_{n\bm{k}}^{\sigma\ast}(\bm{r}) \psi_{m\bm{q}}^{\sigma\ast}(\bm{r'}) \psi_{n\bm{k}}^{\sigma}(\bm{r'}) \psi_{m\bm{q}}^{\sigma}(\bm{r})
    }{|\bm{r}-\bm{r'}|} \erfc(\omega|\bm r - \bm {r'}|)   d\bm{r'} d\bm{r}  \,,
\end{align}
where $\omega$ is the screening parameter that determines the range separation. The corresponding potential appearing in the Hamiltonian, which can be interpreted as the short-range part of the exact exchange operator, takes the form:
\begin{align} \label{Eq:VxSR}
    V_{X,SR}^{\sigma}&\varphi_{n\bm k}^{\sigma}(\bm r) = -\sum_{m\bm q} w_{\bm q} g_{m\bm q}^{\sigma} \psi_{m\bm q}^{\sigma}(\bm r) \nonumber \\
    &\times \int  \frac{\psi_{m\bm q}^{\sigma\ast}(\bm{r'}) \varphi_{n\bm k}^{\sigma}(\bm{r'}) \erfc({\omega |\bm r - \bm{r'}|}) }{|\bm{r}-\bm{r'}|} d\bm{r'} \,,
\end{align}
where $\varphi$ is any given function. Analogous to the unscreened case, the short range part of the exact exchange operator in Eq.~\ref{Eq:VxSR} and the corresponding energy in Eq.~\ref{Eq:ExSR} can be rewritten as: 
\begin{align}
V_{X,SR}^{\sigma}\varphi_{n\bm k}^{\sigma}(\bm r) &= -\sum_{m\bm q} w_{\bm q} g_{m\bm q}^{\sigma} \psi_{m\bm q}^{,\sigma}(\bm r)  \hat{\phi}_{m\bm q n\bm k}^{\sigma} (\bm{r}) \,, \label{Eq:VxSR:Poisson} \\
 E_{X,SR} & = -\frac{1}{2} \sum_{\sigma} \sum_{n\bm{k},m\bm{q}} w_{\bm{k}} w_{\bm{q}} g_{n\bm{k}}^{\sigma} g_{m\bm{q}}^{\sigma}  \nonumber \\ 
& \hspace{6mm} \times \int \psi_{n\bm{k}}^{\sigma\ast}(\bm{r}) \hat{\phi}_{m\bm q n\bm k}^{\sigma} (\mathbf{r})  \psi_{m\bm{q}}^{\sigma}(\bm{r}) dr \,, \label{Eq:ExSR:Poisson} 
\end{align}
where $\hat{\phi}$ represents the solution to the equations:
\begin{align} \label{Eq:Poisson:SR}
-\frac{1}{4 \pi} \left( I - e^{ -\frac{\nabla^2}{16 \pi \omega^2} } \right)^{-1}  \nabla^2 \hat{\phi}_{m\bm q n\bm k}^{\sigma}(\bm{r}) = \psi_{m\bm q}^{\sigma\ast}(\bm{r}) \varphi_{n\bm k}^{\sigma}(\bm{r}) \,,
\end{align}
again subject to Bloch boundary conditions at  the  wavevector $\bm k - \bm q$ in the directions that the system is extended, and  Dirichlet boundary conditions in the  directions of vacuum. To arrive at  Eq.~\ref{Eq:Poisson:SR}, we derived that $\frac{\erfc({\omega |\bm r - \bm{r'}|})}{|\bm{r} - \bm{r}'|}$ is the Green's function of  the $-\frac{1}{4 \pi} \left( I - e^{ -\frac{\nabla^2}{16 \pi \omega^2} } \right)^{-1}  \nabla^2$ operator. Here and above, $I$ denotes the identity operator.  The expressions for isolated systems can again be obtained by setting the wavevectors $\bm k = \bm q=0$ and the Brillouin zone weight $\bm{w}_0=1$.


\section{Real space formulation} \label{Sec:Formulation}

In calculations involving hybrid exchange-correlation functionals, the computational cost is  determined primarily by the solution of the associated linear systems, i.e., Eqs.~\ref{Eq:Poisson} and \ref{Eq:Poisson:SR} for the unscreened and range-separated variants, respectively. In particular,  $\mathcal{O}(N_k^2 N_s^2)$ linear systems need to solved in each simulation, where $N_k$ is the number of wavevectors, and $N_s$ is the number of occupied states. We now develop a highly efficient Kronecker product based real space formalism for the solution of these linear systems, as described below.

The Laplace operator takes the following form in the Cartesian coordinate system:
\begin{align}\label{eq:Laplacian}
\nabla^2 =  \frac{\partial^2}{\partial x_1^2}+\frac{\partial^2}{\partial x_2^2}+\frac{\partial^2}{\partial x_3^2} \,,
\end{align}
where $(x_1, x_2, x_3)$  represent the coordinates. Consider a uniform 3D grid of $N = n_1 n_2 n_3$ points, with $n_1$, $n_2$, and $n_3$ grid points along the $x_1$, $x_2$, and $x_3$ directions, respectively. The discrete sparse Laplacian matrix within the centered finite-difference approximation admits the following decomposition \cite{sharma2018real}:
\begin{align}
\bL  &=   \bIz \otimes \bIy \otimes \bLx  + \bIz \otimes \bLy \otimes \bIx  + \bLz \otimes \bIy \otimes \bIx 
\end{align} 
where $\otimes$ denotes the Kronecker product \cite{van2000ubiquitous, golub2013matrix}; $\bIx$, $\bIy$, and $\bIz$ are the identity matrices of size $n_1$, $n_2$, and $n_3$, respectively; and $\bLx$, $\bLy$, and $\bLz$ are the centered finite-difference approximation of the $\frac{\partial^2}{\partial x_1^2}$, $\frac{\partial^2}{\partial x_2^2}$, and $\frac{\partial^2}{\partial x_3^2}$ operators, respectively. In particular, the matrices $\bLx$, $\bLy$, and $\bLz$ correspond to the finite-difference approximation of the second deriviative operators on the uniform 1D grids that comprise the 3D grid in the $x_1$, $x_2$, and $x_3$ directions, respectively. Therefore, they are square matrices of sizes $n_1$, $n_2$, and $n_3$, respectively, and incorporate the prescribed boundary conditions, i.e., Dirichlet, periodic, or Bloch-periodic,  along the $x_1$, $x_2$, and $x_3$ directions, respectively, as formulated within the real-space method \cite{ghosh2017sparc, GHOSH2017109}.  Though the matrices can  indeed be wavevector dependent, we do not index it as such here for notational simplicity and clarity of discussion. 

Consider the eigendecomposition of the  finite-difference matrices $\bLx$, $\bLy$, and $\bLz$: 
\begin{align}
\bLx {\boldsymbol \Lambda}_1 = \bVx {\boldsymbol \Lambda}_1 \,, \,\, \bLy {\boldsymbol \Lambda}_2 = \bVy {\boldsymbol \Lambda}_2 \,, \,\, \bLz {\boldsymbol \Lambda}_3 = \bVz {\boldsymbol \Lambda}_3 \,, 
\end{align}
 where ${\boldsymbol \Lambda}_1$, ${\boldsymbol \Lambda}_2$, and ${\boldsymbol \Lambda}_3$ are diagonal matrices containing the eigenvalues  of $\bLx$, $\bLy$, and $\bLz$, respectively; and $\bVx$, $\bVy$, and $\bVz$ are the eigenvectors of $\bLx $, $\bLy$, and $ \bLz$, respectively.  It follows from the mixed product property \cite{van2000ubiquitous, golub2013matrix} of Kronecker products  that: 
\begin{align} \label{Eq:EigDecomp:L}
\bL \mathbf{V} = \mathbf{V} \boldsymbol{\Lambda} \,,
\end{align}
where 
\begin{align}
& \mathbf{V}  = \bVx \otimes \bVy \otimes \bVz \,, \\
&\boldsymbol{\Lambda} =   \bIz \otimes \bIy \otimes {\boldsymbol{\Lambda}}_1 +  \bIz \otimes {\boldsymbol{\Lambda}}_2 \otimes \bIx  + {\boldsymbol{\Lambda}}_3 \otimes \bIy \otimes \bIx \,.
\end{align}
Since $\boldsymbol{\Lambda}$ is a diagonal matrix, it follows that $\mathbf{V}$ is a matrix of the eigenvectors of $\bL$, and $\boldsymbol{\Lambda}$ is a matrix of the corresponding eigenvalues. Therefore, for any given function: 
\begin{align} \label{Eq:func:L}
f(\bL)  = f(\mathbf{V} \boldsymbol{\Lambda} \mathbf{V}^{-1} ) = \mathbf{V} f(\boldsymbol{\Lambda}) \mathbf{V}^{-1}   \,,
\end{align}
where 
\begin{align}
f(\boldsymbol{\Lambda}) = &  \bIz \otimes \bIy \otimes f({\boldsymbol{\Lambda}}_1) +  \bIz \otimes f({\boldsymbol{\Lambda}}_2) \otimes \bIx  \nonumber  \\ 
 & +  f({\boldsymbol{\Lambda}}_3) \otimes \bIy \otimes \bIx \,.
\end{align}
In Eq.~\ref{Eq:func:L}, the first equality is obtained by right multiplying Eq.~\ref{Eq:EigDecomp:L} with $\mathbf{V}^{-1}$, and the second equality is obtained  using the spectral theorem. Since, $\mathbf{V}$ is an orthogonal matrix, i.e., $\mathbf{V}^{-1} = \mathbf{V}^{* \rm T}$, we arrive at the relation:
\begin{align} \label{Eq:FinalRepfL}
f(\bL)  = (\bVx \otimes \bVy \otimes \bVz)  f(\boldsymbol{\Lambda})  (\bVx^{* \rm T} \otimes \bVy^{* \rm T} \otimes \bVz^{* \rm T}) \,. 
\end{align}

The solution to the linear systems of interest  can be written as:
\begin{align} \label{Eq:LS:Discrete:Solution}
\bX = f(\bL) \bB \,,
\end{align}
where $\bB$ denotes the right hand side vector, $f(\bL) = -4 \pi \bL^{-1}$ for Eq.~\ref{Eq:Poisson}, and $ f(\bL) = -4 \pi \bL^{-1} (\mathbf{I} - e^{-\frac{\bL}{16 \pi \omega^2}})$ for Eq.~\ref{Eq:Poisson:SR}. Though it is indeed possible to calculate $f(\bL)$ using  Eq.~\ref{Eq:FinalRepfL}, a strategy that circumvents the need for the direct eigendecomposition of $\bL$, it is not an attractive one since the resulting matrix will be a large dense matrix  that is problematic to store. In addition, the solution of each linear system will scale quadratically with the total number of grid points, i.e., $\mathcal{O}(N^2)$. In view of this, we use Roth's relationship \cite{roth1934} to write the solution as:
\begin{subequations} 
\label{Eq:FinalSolution}
\begin{align} 
& \bX = {\rm vec}_{n_3} \left[ \left( \bigwedge_{1 \le k\le n_3}  {\rm vec}_{n_2} (\bVx \widetilde{\bX}_k \bVy^{\rm T})  \right) \bVz^{\rm T} \right] \,,  \\
& \widetilde{\bX} = f(\boldsymbol{\Lambda}) \odot  {\rm vec}_{n_3} \left[ \left( \bigwedge_{1 \le k\le n_3}  {\rm vec}_{n_2} (\bVx^{* \rm T} \bB_k \bVy^{*}) \right) \bVz^{*} \right] \,,
\end{align}
\end{subequations}
where ${\rm vec}_{(.)}$ denote the vectorization operator  along the subscripted  dimension, which converts the matrix to a column vector \cite{van2000ubiquitous, sharma2018real}; $\bigwedge$ represents the loop operator, which accumulates the different column vectors into a matrix \cite{sharma2018real}; $\widetilde{\bX}_k = \bX(:,:,k)$ and $\bB_k = \bB(:,:,k)$ denote the frontal slices of the $\bX$ and $\bB$ vectors, respectively, while within a multidimensional representation; and $\odot$ represents the Hadamard (element wise) product.  

\begin{figure}[htbp]
\centering
  \includegraphics[width=0.4\textwidth]{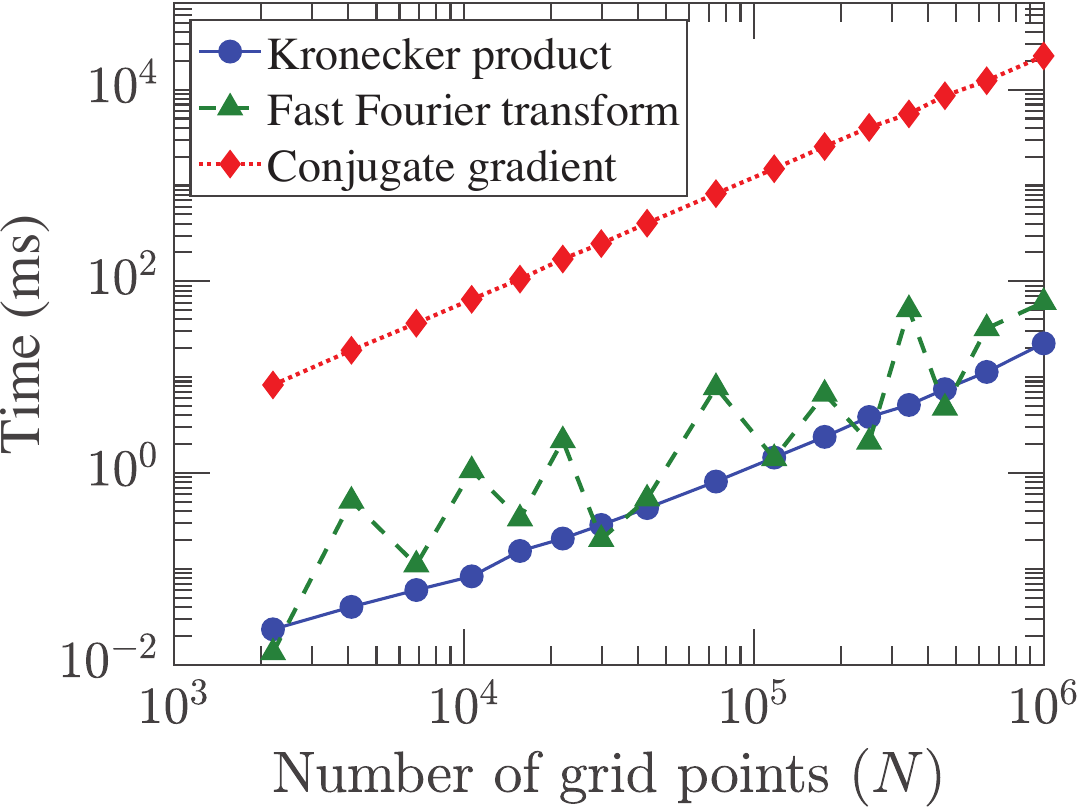}
  \caption{Time to solution for the Poisson equation as a function of the number of grid points ($N$) for the Kronecker product, fast Fourier transform (FFT), and conjugate gradient methods. The domain is cubical and has $N^{1/3}$ grid points in each direction. Larger values of $N$ is achieved by increasing the domain size, while holding the grid spacing fixed. Routines available within the Math Kernel Library (MKL) have been used for performing the FFTs.  The initial guess and stopping tolerance (in the relative residual) for the conjugate gradient method are a vector of all ones and $10^{-3}$, respectively. The jagged nature of the curve for the solution via FFT can be attributed to variations in the speed of FFT when the number of grid points in each direction is not a power of two. In all instances, the timings are averaged over 30 runs. \label{Fig:compareFFTCG}}
\end{figure}

The computational cost associated with the solution of each linear system  scales as: $\mathcal{O}(2n_1^2n_2) + \mathcal{O}(2n_1n_2^2) + \mathcal{O}(2 n_1 n_2 n_3^2) + \mathcal{O}(n_1 n_2 n_3) \sim \mathcal{O}(n_1 n_2 n_3^2)$.   Therefore, depending on the number of grid points in each direction, the overall scaling is between $\mathcal{O}(N)$ and $\mathcal{O}(N^{4/3})$. Indeed, if $n_3 \gg n_2 \sim n_1$, the axes can be interchanged to ensure that the scaling is reduced from $\mathcal{O}(N^2)$ to $\mathcal{O}(N)$. In electronic structure calculations it is typical for $n_1 \sim n_2 \sim n_3 \sim N$, whereby the computational cost scales as $\mathcal{O}(N^{4/3})$ with the number of grid points. Note that in the case of periodic/Bloch-periodic boundary conditions in all directions, the matrix products involving $\bVx$, $\bVy$, and $\bVz$ can all be evaluated using FFTs, whereby the overall scaling  will reduce to $\mathcal{O}(N \log N $), the same as when FFTs are directly used for the 3D problem. Importantly, even with the $\mathcal{O}(N^{4/3})$ scaling, the Kronecker product based formalism is competive with 3D FFTs, as demonstrated in Fig~\ref{Fig:compareFFTCG}, while not being restricted to periodic/Bloch-periodic boundary conditions. Also note that iterative solution techniques for the solution of linear systems can formally scale as  $\mathcal{O}(N)$, however, the associated prefactor can be up to three orders of magnitude larger, as shown in Fig.~\ref{Fig:compareFFTCG}. Furthermore, such a scheme is only applicable to Eq.~\ref{Eq:Poisson}, while being impractical for the solution of Eq.~\ref{Eq:Poisson:SR}, since it requires calculating the exponential of $\bL$, which is a large sparse matrix whose size ranges from thousands to tens of thousands times the number of atoms in the system \cite{xu2018discrete}. 

The formalism presented above can be easily extended to non-orthogonal cells for periodic/Bloch-periodic boundary conditions, i.e., 3D bulk systems. In particular, starting with the generalized form of the Laplacian that also has second-order mixed derivatives, and utilizing the fact that the 1D first and second derivative finite-difference matrices share a common eigenbasis, the above procedure can be utilized to derive the solution, which bears significant similarity  in form to the orthogonal case presented above. However, this strategy is  not applicable to non-orthogonal cells when even one of the directions has Dirichlet boundary conditions, e.g., slabs, since the 1D first and second derivative finite-difference matrices no longer commute. 


\section{Implementation} \label{Sec:Implementation}

We have implemented the aforedescribed Kronecker product based real space formalism for hybrid density functionals in the SPARC electronic structure code \cite{zhang2024sparc, xu2021sparc}, after having developed a prototype implementation in its MATLAB version M-SPARC \cite{zhang2023version, xu2020m}. In particular, building on the existing implementation for the Perdew–Burke–Ernzerhof (PBE) functional \cite{perdew1996generalized}, we implement the PBE0 \cite{Becke1993DensityfunctionalTI, perdew1996rationale} and Heyd–Scuseria–Ernzerhof (HSE) \cite{heyd2003hybrid, krukau2006influence} hybrid functionals, wherein the exchange-correlation energy takes the form:
\begin{align}
    E_{xc}^{PBE0} & =\frac{1}{4}E_X+\frac{3}{4}E_{x}^{PBE}+E_c^{PBE} \,, \\
    E_{xc}^{HSE} &= \frac{1}{4}E_{X,SR} + \frac{3}{4}E_{x,SR}^{PBE} + E_{x,LR}^{PBE} + E_c^{PBE} \,.
\end{align}
Above,  $E_{x}^{PBE} = E_{x,SR}^{PBE} + E_{x,LR}^{PBE}$ and $E_c^{PBE}$ are the PBE exchange and correlation energies, respectively, with $E_{x,SR}^{PBE}$ and  $E_{x,LR}^{PBE}$ representing the short range and long range components of the PBE exchange energy. 

In each simulation, we first calculate the electronic ground state corresponding to the PBE exchange-correlation functional, with the orbitals and density so determined used as an initial guess for the hybrid functional calculation. We employ an outer fixed-point iteration with respect to the exact exchange operator \cite{lin2016adaptively} (the corresponding exact exchange energy used to decide convergence), in addition to the usual inner fixed-point iteration with respect to the density/potential, commonly referred to as the self-consistent field (SCF) method \cite{martin2020electronic}. In particular, we employ the adaptively compressed exchange (ACE) operator  method \cite{lin2016adaptively} in conjunction with the existing SPARC implementation of the Chebyshev filtered subspace iteration (CheFSI) \cite{zhou2006parallel, zhou2006self}, which is accelerated using the restarted variant of the preconditioned Periodic Pulay mixing scheme \cite{banerjee2016periodic, pratapa2015restarted, kumar2020preconditioning}, with the Poisson problem for the electrostatics \cite{ghosh2016higher, suryanarayana2014augmented} solved using the alternating Anderson-Richardson (AAR) method \cite{suryanarayana2019alternating, pratapa2016anderson}. In each SCF iteration, a four level parallelization scheme is adopted for the application of the ACE operator, i.e., over spin, Brillouin zone wavevectors, band, and domain, in that order, as implemented in SPARC for local/semilocal exchange-correlation functionals \cite{xu2021sparc}. 

In setting up the ACE operator, the linear systems, i.e., Eqs.~\ref{Eq:Poisson} and \ref{Eq:Poisson:SR} for the PBE0 and HSE functionals, respectively, are solved using the Kronecker product based formalism described in the previous section (Eq~\ref{Eq:FinalSolution}). For isolated systems, the linear systems are subject to Dirichlet boundary conditions, determined through the multipole expansion of the Coulomb kernel,  which are enforced in the real-space method through suitable modification of the right hand side vector \cite{ghosh2017sparc}.  For solid state systems, the linear systems are subject to periodic/Bloch-periodic boundary conditions in the directions that the system is extended, and Dirichlet boundary conditions in the directions that the system has vacuum, if any. In the case of 3D bulk systems and  PBE0 exchange-correlation functional, there is a singularity present in the Poisson problem (Eq.~\ref{Eq:Poisson}) for $\bm k = \bm q$, which is overcome through auxiliary functions \cite{gygi1986self, wenzien1995efficient, carrier2007general}. 

Each linear system appearing in the construction of the ACE operator is solved serially, with the right hand side vectors obtained by first performing cyclic communication among all the band communicators to gather the local part of the orbitals from the neighboring communicator to the left, then constructing the local parts of the right hand side vectors, and finally, performing a communication between all the domain communicators within each band communicator using the \texttt{MPI\_Alltoallv} routine. Once the solution has been obtained, the solution is transferred back to the original layout, again using the \texttt{MPI\_Alltoallv} routine.  This procedure is repeated until all the linear systems have been solved. Indeed, in the presence of Brillouin zone integration and spin, this procedure is adopted for the band and domain communicators within each wavevector communicator, all of which are within each spin communicator. During the solution of each linear system, the matrix-matrix products appearing in Eq~\ref{Eq:FinalSolution} are performed using Level 3 BLAS routines.


\section{Results and discussion} \label{Sec:Results}

In this section, we first verify the accuracy and performance of the Kronecker product based real-space formalism for hybrid density functionals, and then use it to study the  structure of liquid water, as determined from AIMD. 

\subsection{Accuracy and performance} \label{Subsec:AccPerf}
We consider the following three systems: 247-atom hydrogen terminated silicon nanocluster (Si$_{147}$H$_{100}$), perturbed 512-atom cell of bulk lithium hydride (LiH), and  perturbed 36-atom cell of bulk manganese dioxide (MnO$_2$). In particular, we employ the PBE0 exchange-correlation functional for Si$_{147}$H$_{100}$, and the HSE06 exchange-correlation functional for the LiH and MnO$_2$ systems. In all instances, we employ optimized norm conserving Vanderbilt (ONCV) pseudopotentials \cite{hamann2013optimized} from the SPMS set \cite{shojaei2023soft}, which have been  developed for the PBE exchange-correlation functional with nonlinear core corrections (NLCC), having 4, 1, 3, 15, and 6 electrons in valence for the Si, H, Li, Mn, and O chemical elements, respectively. We perform spin-unpolarized calculations for the Si$_{147}$H$_{100}$ and LiH systems, and spin-polarized calculations for the MnO$_2$ system. We perform Brillouin zone integration using the $\Gamma$-point and $2 \times 2 \times 2$ grid for the LiH and MnO$_2$ systems, respectively. 

In SPARC, we use real-space grid spacings of 0.33, 0.32, and 0.26 bohr for the  Si$_{147}$H$_{100}$, LiH, and MnO$_2$ systems, respectively, whereby the energies and atomic forces are converged to within $10^{-3}$ ha/atom and $10^{-3}$ ha/bohr, respectively. We compare our results and timings with the established planewave code Quantum Espresso \cite{giannozzi2017advanced, Espresso, barnes2017improved} (QE) v7.0. In QE, just as in SPARC, the PBE solution is used as the initial guess, and we select the option of the ACE method, which makes QE consistent with the SPARC implementation. Moreover, we choose planewave energy cutoffs of 25, 25, and 35 ha for the Si$_{147}$H$_{100}$, LiH, and MnO$_2$ systems, respectively, which ensures that the resulting accuracies are comparable to those in SPARC. To facilitate comparison, the vacuum  for the Si$_{147}$H$_{100}$ system is set to 8.5 bohr in both codes, though we note that SPARC will generally require smaller amount of vacuum, a consequence of the ability to incorporate Dirichlet boundary conditions.  The default parallelization settings are employed in SPARC, whereas they are optimized in QE, given that QE can be made to be up to an order-of-magnitude faster in time to solution relative to the default parallelization settings. The stopping criterion and tolerance for the outer loop in SPARC is set to be the same as that in QE. All other parameters are set to their default values in both codes. 

\begin{figure*}[htbp]
  \centering
  \subfloat[Si$_{147}$H$_{100}$]{
    \includegraphics[width=0.32\textwidth]{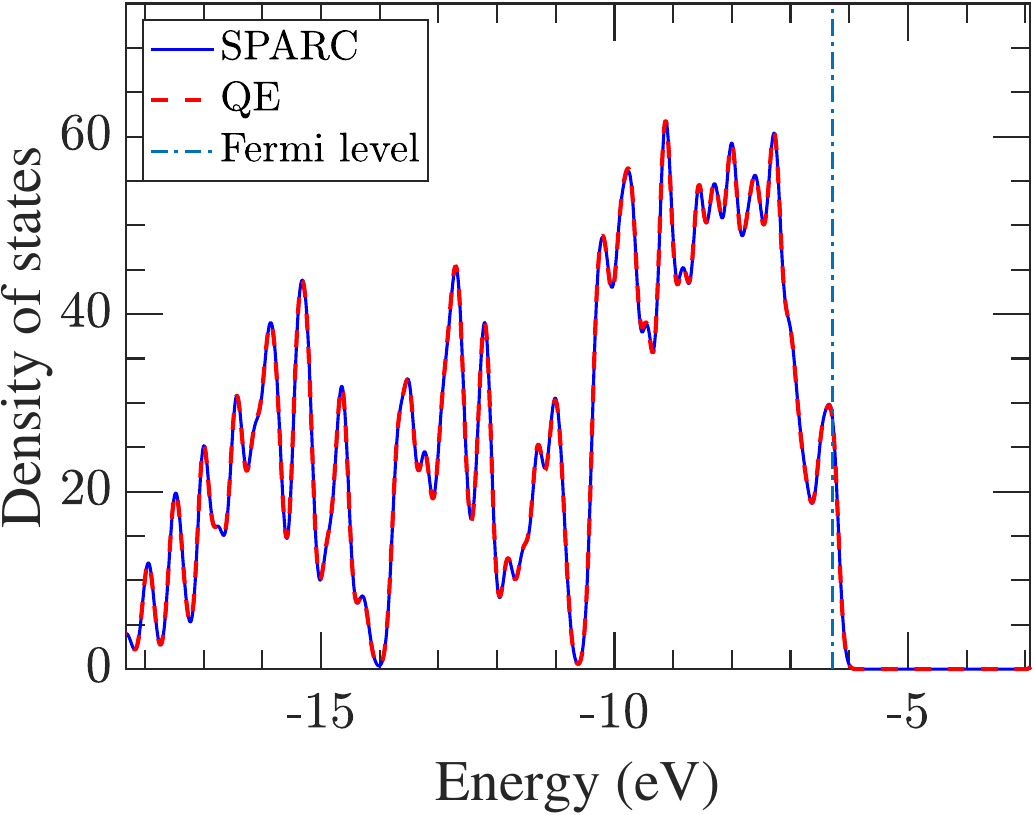}
    \label{dos-SiH}}
     \subfloat[LiH]{
    \includegraphics[width=0.32\textwidth]{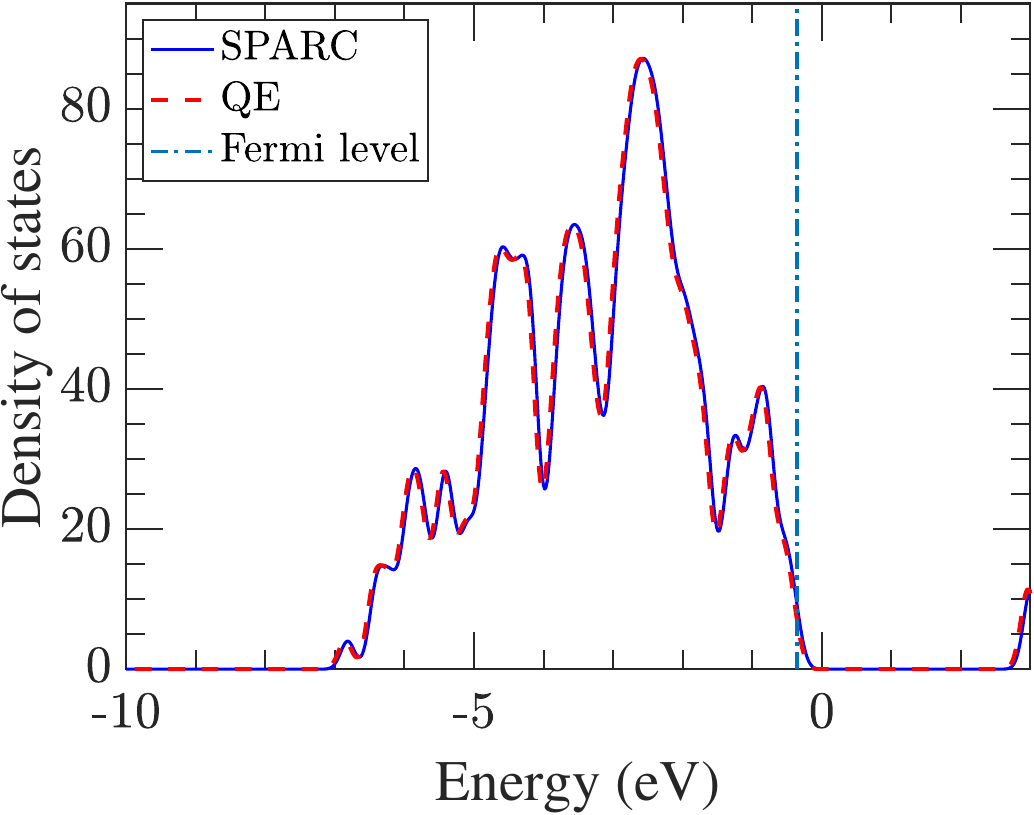}
    \label{dos-LiH}}
      \subfloat[MnO$_2$]{
    \includegraphics[width=0.32\textwidth]{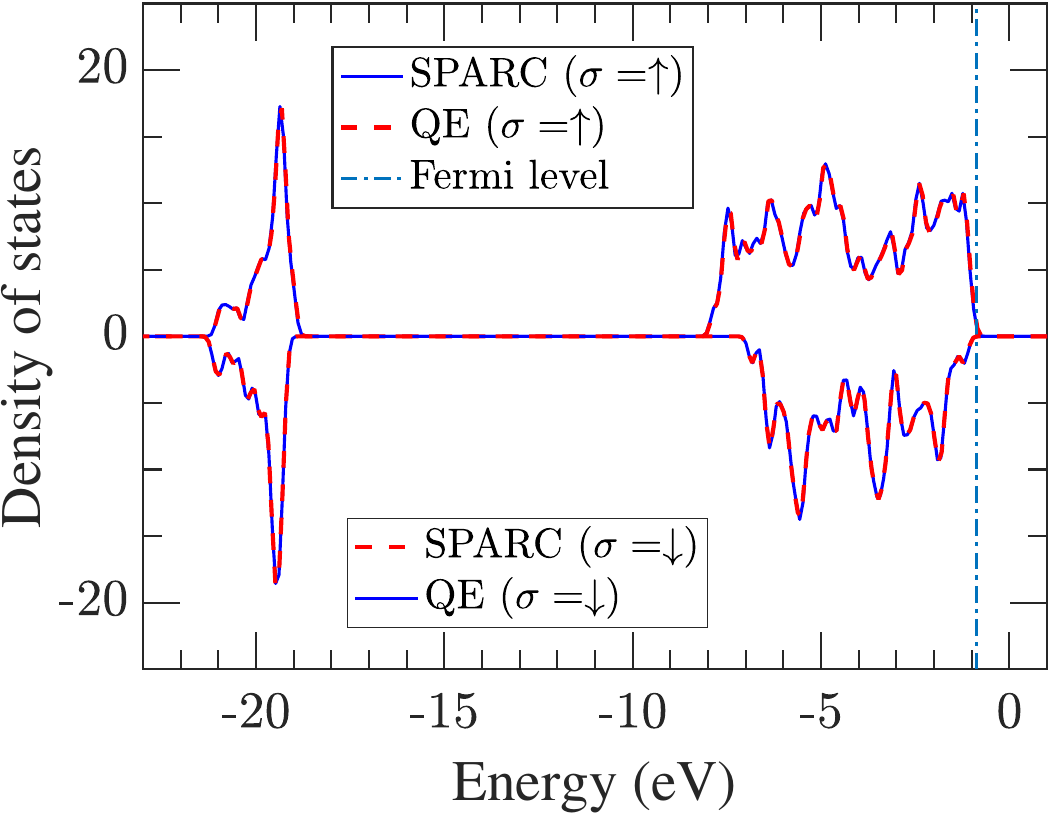}
    \label{dos-MnO}} 
    \caption{\label{Fig:DOS} Density of states  (DOS) for the Si$_{147}$H$_{100}$, LiH, and MnO$_2$ systems, as computed by QE and the  Kronecker product based real space formalism in SPARC.}
  \end{figure*}

In Fig.~\ref{Fig:DOS}, we compare the density of states (DOS) computed by SPARC and QE for the Si$_{147}$H$_{100}$, LiH, and MnO$_2$ systems. For plotting the DOS, we choose a Gaussian broadening parameter of $0.1$ eV. We observe that there is excellent agreement between SPARC and QE for all the systems, with a maximum difference in the eigenvalues of $2 \times 10^{-4}$, $7 \times 10^{-4}$, and $10^{-3}$ ha  for the Si$_{147}$H$_{100}$, LiH, and MnO$_2$ systems, respectively. Indeed, there is also very good agreement in the energies and forces computed by SPARC and QE, with a difference in the energy of $9 \times 10^{-5}$, $5 \times 10^{-4}$, and $7 \times 10^{-4}$ ha/atom, and a maximum difference in any force component over all the atoms of $8 \times 10^{-4}$, $8 \times 10^{-4}$, and $10^{-3}$ ha/bohr for the Si$_{147}$H$_{100}$, LiH, and MnO$_2$ systems, respectively. The agreement further increases on choosing larger number of grid points and planewave energy cutoffs in SPARC and QE, respectively. For instance, on choosing a grid spacing of $0.2$ bohr in SPARC and a planewave energy cutoff  of $50$ ha in  QE for the LiH system, the eigenvalues, energy, and forces  agree to within $3 \times 10^{-4}$ ha, $9 \times 10^{-5}$ ha/atom, and $3 \times 10^{-4}$ ha/bohr, respectively.

\begin{figure*}[htbp]
  \centering
  \subfloat[Si$_{147}$H$_{100}$]{
    \includegraphics[width=0.32\textwidth]{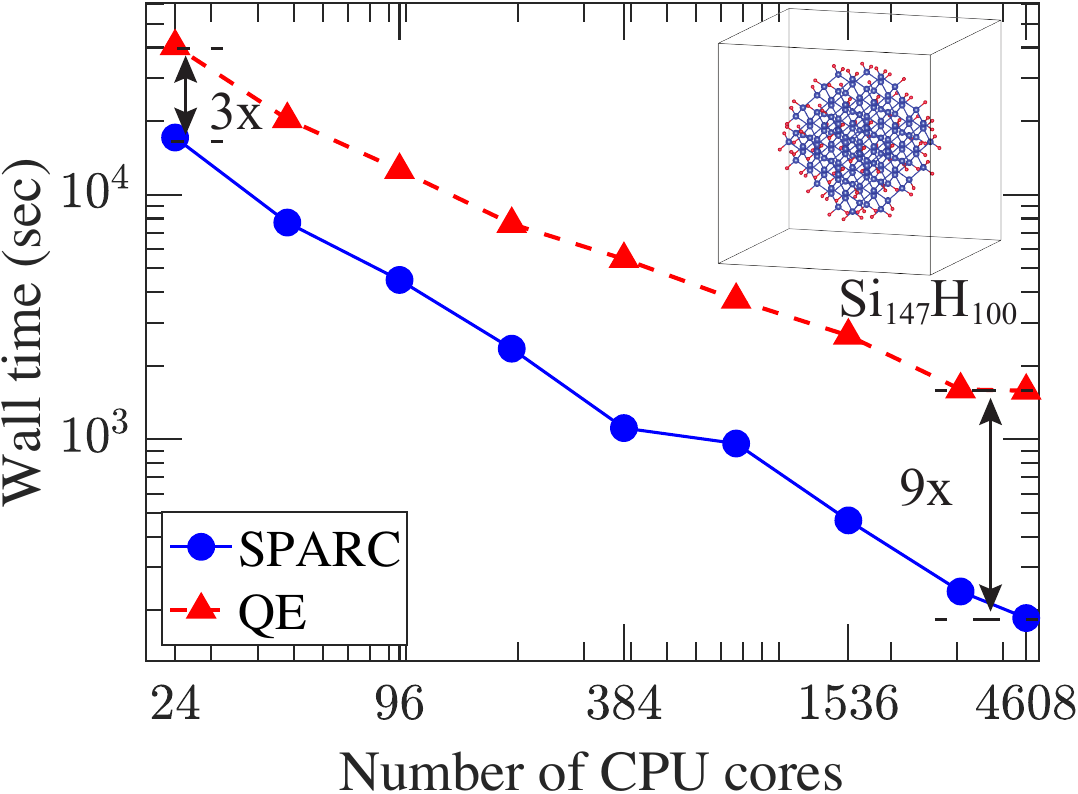}
    \label{ss-SiH}}
     \subfloat[LiH]{
    \includegraphics[width=0.32\textwidth]{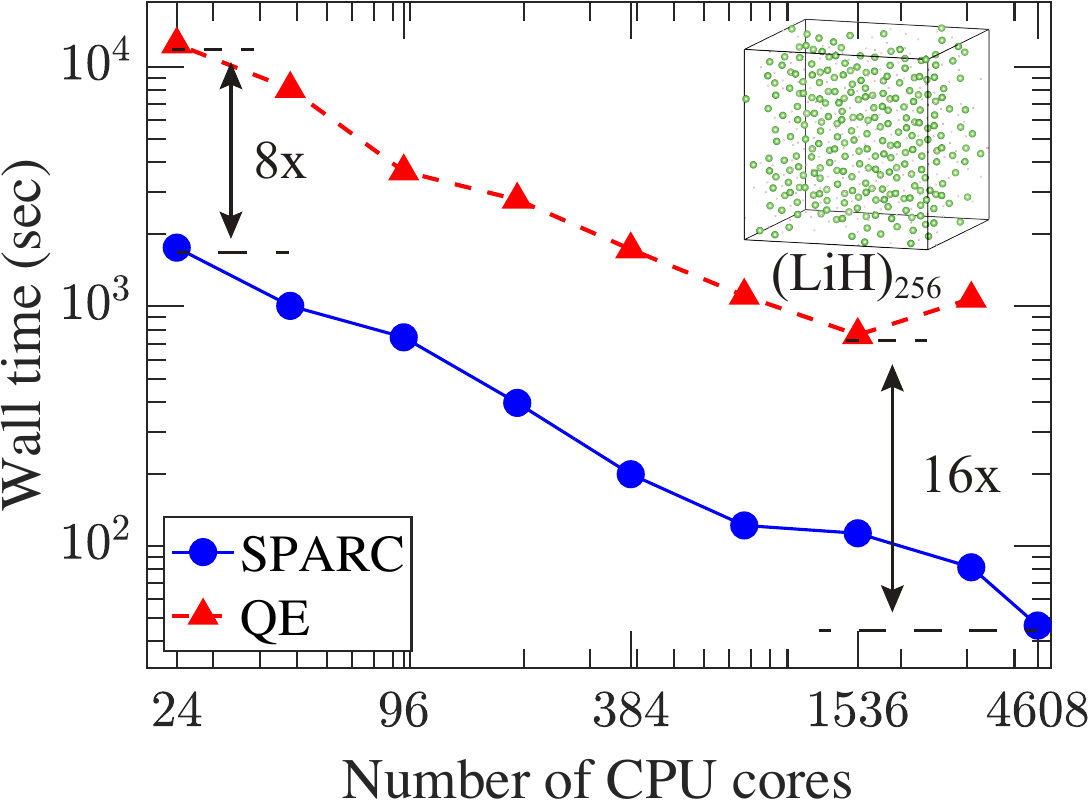}
    \label{ss-LiH}}
      \subfloat[MnO$_2$]{
    \includegraphics[width=0.32\textwidth]{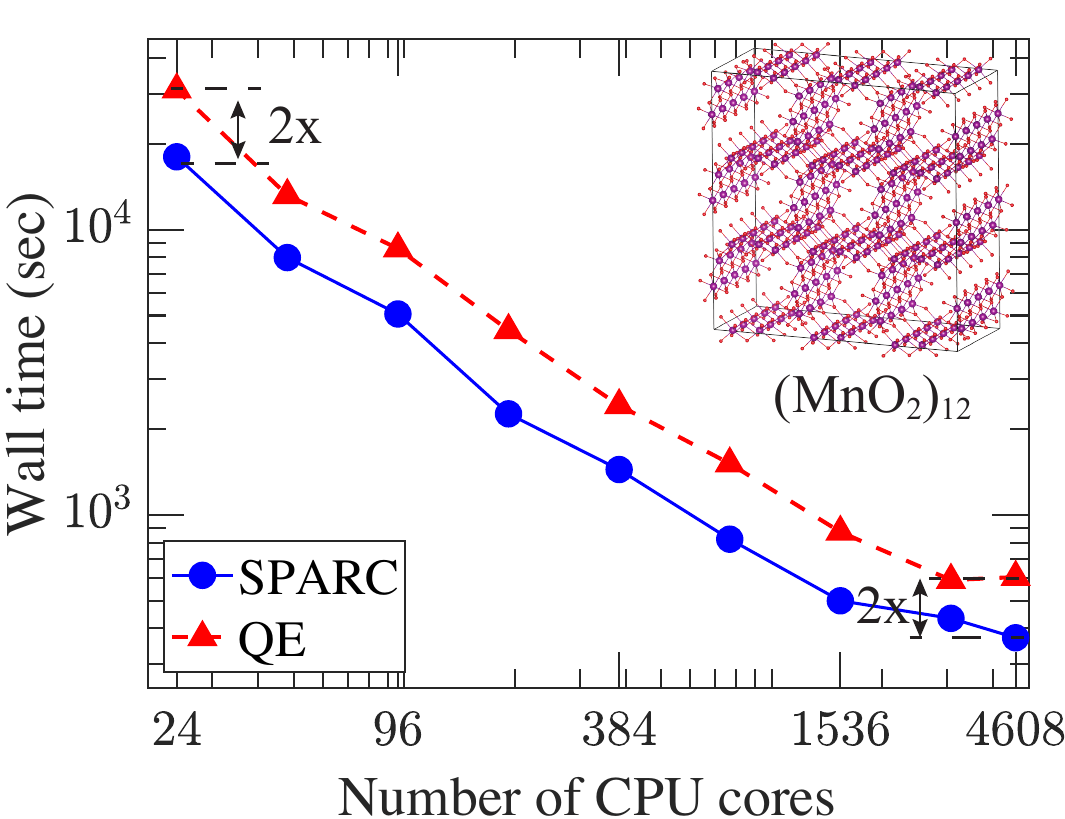}
    \label{ss-MnO}} 
    \caption{\label{Fig:StrongScaling} Strong scaling of QE and the Kronecker product based real space  formalism in SPARC for the Si$_{147}$H$_{100}$, LiH, and MnO$_2$ systems.}
  \end{figure*}

In Fig.~\ref{Fig:StrongScaling}, we compare the strong scaling performance of SPARC and QE for the Si$_{147}$H$_{100}$, LiH, and MnO$_2$ systems. In particular, we present the variation in wall time as the number of CPU cores is increased from 24 to 4608. We find that SPARC scales to  thousands of processors even though the systems under consideration are of moderate size. This is indeed true also for QE, however the minimum time to solution in SPARC is smaller by factors of $9$, $16$, and $2$ for the Si$_{147}$H$_{100}$, LiH, and MnO$_2$ systems, respectively. Furthermore, even on 24 CPU cores, the time to solution in SPARC  is smaller by factors of  $3$, $8$, and $2$ for the Si$_{147}$H$_{100}$, LiH, and MnO$_2$ systems, respectively. These speedups can be attributed to the increased efficiency of SPARC during the application of the ACE operator, a consequence of domain parallelization being more efficient in real-space methods compared to planewave methods, by virtue of the locality of the real-space representation relative to the Fourier representation. Indeed, in its default settings for hybrid functional calculations, SPARC adopts a much higher preference towards domain parallelization relative to band parallelization, the opposite being true for local/semilocal exchange-correlation functionals. Given that the Si$_{147}$H$_{100}$ and LiH systems have substantially more grid points relative to the MnO$_2$ system, the potential for domain parallelization is significantly higher in Si$_{147}$H$_{100}$ and LiH, enabling larger speedups. An even larger speedup would have been achieved for Si$_{147}$H$_{100}$, as is to be expected based on having more than four times the number of grid points as in the LiH system, but for the time spent in the evaluation of the Dirichlet boundary conditions through the multipole expansion.  

As demonstrated in Fig.~\ref{Fig:compareFFTCG}, the Kronecker product formalism can outperform iterative solvers by  up to three orders of magnitude for the solution of a single linear system. This translates to more than two orders-of-magnitude speedup in hybrid functional calculations, e.g., considering the Si$_{147}$H$_{100}$ system, the  Kronecker product formalism and conjugate gradient scheme take 123 and 4946 s on 1536 processors, respectively, which represents a 113 factor speedup. 

\subsection{Application: structure of water} \label{Subsec:Water}
We now apply the developed framework to study the structure of liquid water at a density of 0.99656 g/ml and temperature of 330 K, as determined from an AIMD simulation. In particular, we consider 64 water molecules, PBE0 exchange-correlation functional, ONCV pseudopotentials from the SPMS set, and $\Gamma$-point for Brillouin zone integration. We perform isokinetic ensemble (NVK) AIMD simulations with the Gaussian thermostat \cite{minary2003algorithms}  for a total time of 10 ps, with the first 0.2 ps used for equilibration. We use a time step of 0.25 fs, which translates to a total of 40,000 steps, with 800 steps for equlibration. We choose a grid spacing of 0.25 bohr, which converges energies and forces to within $10^{-4}$ ha/atom and $10^{-3}$ ha/bohr, respectively. All other parameters are set to their default values in the SPARC code. 

\begin{figure*}[htbp]
  \centering
  \subfloat[Oxygen - Oxygen]{
    \includegraphics[width=0.32\textwidth]{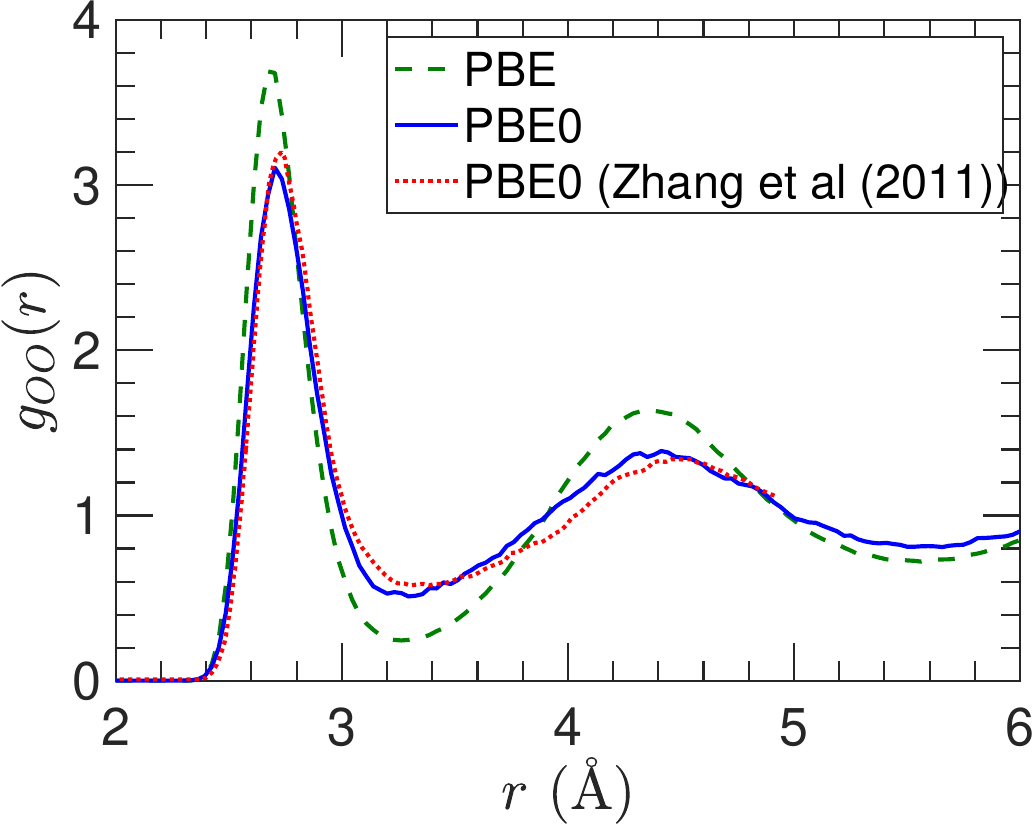}
    \label{rdf-oo}
  }
  \subfloat[Oxygen - Hydrogen]{
    \includegraphics[width=0.32\textwidth]{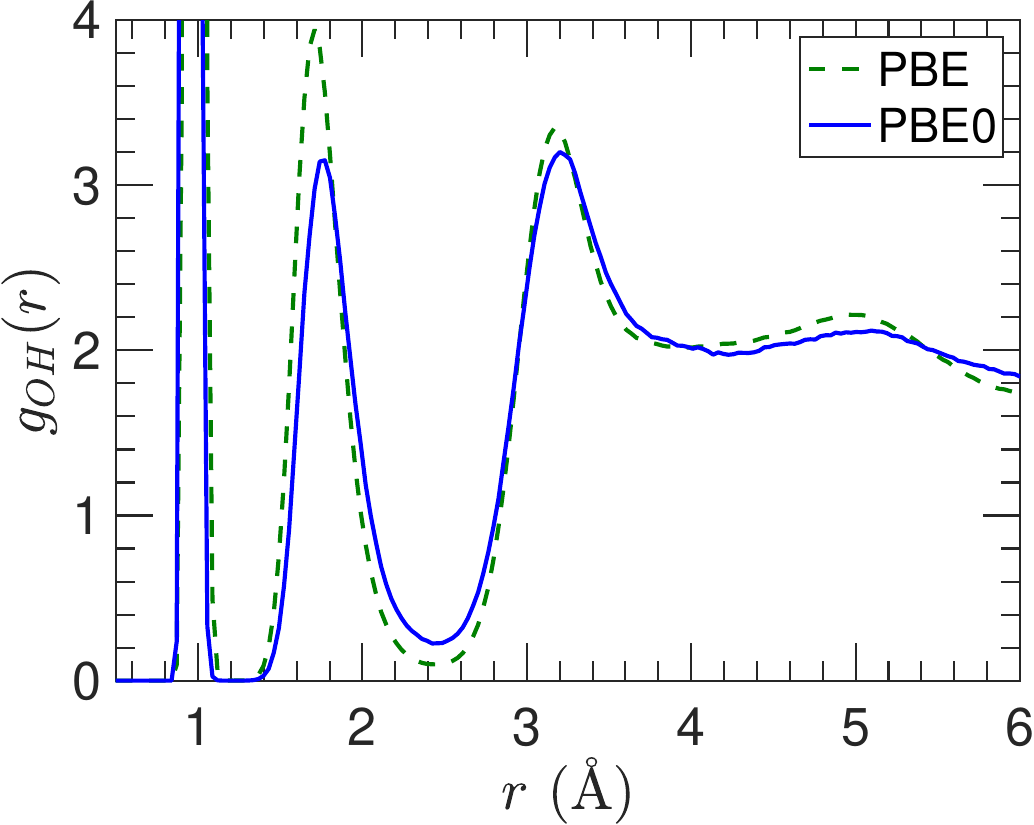}
    \label{rdf-oh}
  }
  \subfloat[Hydrogen - Hydrogen]{
    \includegraphics[width=0.32\textwidth]{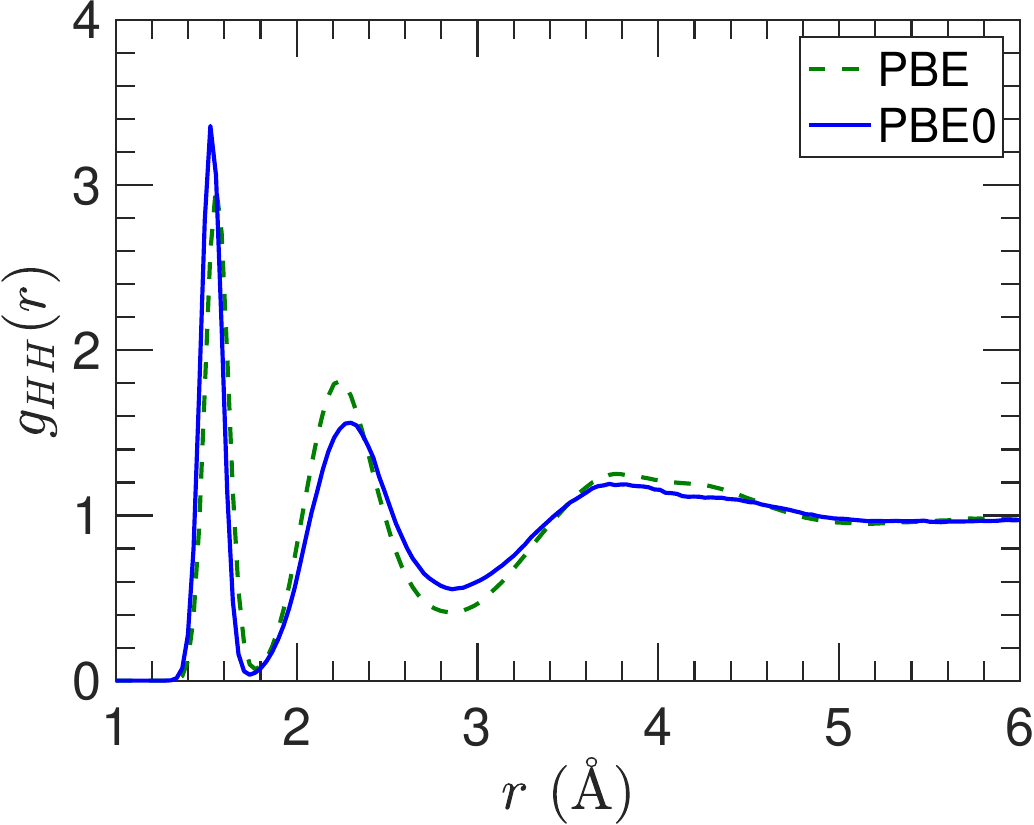}
    \label{rdf-hh}
  }
  \caption{\label{Fig:PDF:water}Pair distribution functions (PDFs) for liquid water at a density of 0.99656 g/ml and temperature of 330 K, as determined from AIMD simulations using the real space Kronecker product based formulation for hybrid functionals implemented in SPARC.}
\end{figure*}

In Fig.~\ref{Fig:PDF:water}, we plot the pair distribution functions (PDFs) for water, i.e., $g_{OO}$, $g_{OH}$, and $g_{HH}$, where the subscript denotes the two elements being considered. For comparison, we plot the $g_{OO}$ available in literature for the PBE0 exchange-correlation functional \cite{zhang2011structural}. We also plot the PDFs obtained for the choice of the PBE exchange-correlation functional, obtained from an AIMD simulation where  everything else is kept the same as the PBE0 simulation. We observe that there is very good agreement between the $g_{OO}$ computed here and that in  literature, verifying the accuracy of the real space formalism. We also observe that there is significant difference in the PDFs obtained for PBE0 and PBE, particularly for the case of $g_{OO}$, which is a well known  limitation of PBE. In particular, PBE0 produces a softer structure than PBE, with the maximum value of $g_{OO}$ decreasing from 3.69 to 3.12, and the minimum value of $g_{OO}$ increasing from 0.24 to 0.52, consistent with the results in literature \cite{zhang2011entropy, zhang2011structural, todorova2006molecular, distasio2014individual}. In terms of performance, the wall time per AIMD step on 480 CPU cores for the PBE0 functional is $\sim$35 seconds, which reduces to $\sim 3$ seconds for the PBE functional. 

\section{Concluding Remarks} \label{Sec:Conclusions}
In this work, we have presented an efficient real space formalism for hybrid exchange-correlation functionals in generalized Kohn-Sham DFT. In particular, exploiting the Kronecker product structure of the real space finite-difference Laplacian matrix, we have developed an efficient representation for any function of this matrix, which enables the solution of  associated linear systems to be competitive with the FFT scheme, without the restriction of periodic/Bloch-periodic boundary conditions. We have implemented this formalism in the SPARC electronic structure code for both unscreened and range-separated hybrid functionals. We have verified the accuracy and efficiency of the  framework by comparing against established planewave codes for isolated as well as bulk systems. Notably, we have demonstrated that the real space framework can achieve up to an order-of-magnitude speedup in time to solution relative to planewave counterparts, with increasing advantages as the number of processors is increased. We have also applied this framework to study the structure of liquid water using AIMD,  the results found to be in good agreement with literature.

The current formalism provides an avenue for efficient real space DFT calculations with hybrid exchange-correlation functionals. It is also more generally applicable to situations where convolutions have to be evaluated with kernels of the form $K(\bm x,\bm y) \equiv K(|\bm x -\bm y|)$. Notably, such situations are regularly encountered in   DFT, including the calculation of the random phase approximation (RPA) correlation energy \cite{Shah2024}, van der Waals density functionals \cite{roman2009efficient}, and preconditioners for the Kohn-Sham eigenproblem \cite{payne1992iterative}, making them  worthy subjects for future research. Analogous to previous work for local/semilocal exchange-correlation functionals \cite{sharma2023gpu}, the GPU implementation of the current formalism is likely to significantly reduce the the time to solution for hybrid functional calculations, making it another worthy subject for future research.

\begin{acknowledgments}
The authors gratefully acknowledge the support of the U.S. Department of Energy, Office of Science under grant DE-SC0023445. This research was also supported by the supercomputing infrastructure provided by Partnership for an Advanced Computing Environment (PACE) through its Hive (U.S. National Science Foundation through grant MRI1828187) and Phoenix clusters at Georgia Institute of Technology, Atlanta, Georgia.
\end{acknowledgments}
\section*{Data Availability Statement}
The data that support the findings of this study are available within the article and from the corresponding author upon reasonable request.
\section*{Author Declarations}
The authors have no conflicts to disclose.

\section*{References}
%


\end{document}